# Automation-driven innovation management?

# Toward Innovation-Automation-Strategy cycle

Piotr Tomasz Makowski[1,*] and Yuya Kajikawa[2,3]

[1] Department of Entrepreneurship and Management Systems, Faculty of Management, University of Warsaw, Szturmowa 1-3, 02-678 Warsaw, Poland
[2] Department of Innovation Science, School of Environment & Society, Tokyo Institute of Technology, Tokyo 108-0023, Japan
[3] Institute for Future Initiatives, The University of Tokyo, 7-3-1 Hongo, Tokyo, 113-8654, Japan
* Corresponding Author: pmakowski@wz.uw.edu.pl (P. Makowski)

## ABSTRACT

There is a resurging interest in automation because of rapid progress of machine learning and AI. In our perspective, innovation is not an exemption from their expansion. This situation gives us an opportunity to reflect on a direction of future innovation studies. In this conceptual paper, we propose a framework of innovation process by exploiting the concept of unit process. Deploying it in the context of automation, we indicate the important aspects of innovation process, i.e. human, organizational, and social factors. We also highlight the cognitive and interactive underpinnings at micro- and macro-levels of the process. We propose to embrace all those factors in what we call Innovation-Automation-Strategy cycle (IAS). Implications of IAS for future research are also put forward.

## 1. INTRODUCTION

Emergent technologies have an impact on existing organizational practice, processes and strategy, and empower or degrade those. They also surge those to transform into *entirely* new forms and also influence end users and society. When management practice and strategy suitably respond to available technologies and fully use them, they contribute to further diffusive development of new technologies in a way that can be highly profitable not only for business and organizations, but also for society.

Especially, contemporary development of AI has tremendous impact on how we understand innovations. Its standard picture, involving highlighted role of innovative and creative individual managers and entrepreneurs has been changing due to the societal changes resulting from the increased use of technology. It has moved automation—the use of technology in various dimensions of human life that makes them more efficient with decreased human assistance—beyond previously known borders. And advancement in the AI is one of the main reasons of this situation.

Although the question of machines thinking and acting like humans is not new (Turing 1950, Lake et al. 2016), it can be perceived as one of the current grand societal challenges (George et al. 2016). Reasoning is no longer the domain of humans (Bottou 2014). Programming is no longer the domain of human programmers (Microsoft's program *DeepCoder* can learn itself how to program (Balog et al. 2016)). Machine learning and deep learning is now being catalyzed by better understanding of how to deal with big data (uncertainty modelling, ensemble learning, etc.) (Sejnowski 2018). Quantum information science (Ladd et al. 2010) is now gaining an experimental realization and claims quantum advantage (Arute et al. 2019). The research on AI has now moved from 'black box' characteristics, entering—as XAI (*Explainable AI*)—the stage of interpretability (Rai 2020, Gill and Hall 2018). Large-scale development, implementation and improved understanding of AI will surely affect the innovation curve (so called *Moore's Law* (Grier 2006)).

Those and other related achievements caused that automation—a broad use of emergent technologies and the AI to make the functioning of business and organizational processes more efficient—became a standard in many areas. AI not only takes routine jobs (Frey and Osborne 2017, Frey 2019), but when it is designed as augmentation or AI-human collaboration (Raisch and Krakowski 2020), it can boost performance by three times in comparison to the abilities of mere computerization (Wilson and Daugherty 2018). Despite some challenges related to AI ethics (Hagendorff 2020, Jobin, Ienca, and Vayena 2019), recent accomplishments in robotics, computer vision, face detection, speech recognition, natural language processing and the like display possibilities of their growing success in tech industry. All these automation-related phenomena change the role of technology in the context of managerial practice, so they also redraw the picture of managing innovations. Thanks to the increased automation, the question of innovation management goes beyond known problems such as: internal vs. external sources of innovation or social pull vs. technology push (Di Stefano, Gambardella, and Verona 2012).

Various technological, organizational and societal phenomena give reasons for the above stipulation—from the expansion of AI and new technologies (machine learning, deep learning, programmable agents, etc.) to the growing awareness of the social role of innovations ('innovation prone' society, sustainable innovation, innovation-driven economic and societal growth). In this light, our concern is how the accumulated expertise in innovation studies can contribute to realization of innovation and accelerate innovation process like or by utilizing emergent AI technologies.

## 2. UNIT PROCESS AND INTEGRATED FRAMEWORK OF INNOVATION PROCESS—TOWARD AUTOMATION OF INNOVATION PROCESS

The aim of this paper is two-fold: we discuss the possibility of automation of innovation process in organization and present a perspective on the role of innovation studies that draws systematic consequences of automation. As for the first, our hypothesis is that given the current success and omnipresence of technology there are reasons to perceive the societal process of innovation as highly automatized. As for the second, to draw actionable consequences of this idea we propose to reconsider the character and role of innovation management and innovation studies in a new perspective. On our view, current knowledge in these areas is developed enough to be examined in what we call the Innovation-Automation-Strategy (IAS) cycle. To obtain suitable context for the IAS cycle, we first introduce integrated framework for the process of innovation.

In their seminal contribution, Frey and Osborne (2017) analyzed the impact of machine learning and robotics on labor market. According to their analysis, bottleneck of computerization of labor includes creative and social intelligence, which is highly related to innovation. Even though they are hard to replace by these emergent technologies, it is helpful and even constructive, we think, to pursue a way where innovation can be replaced or at least supported by those, which means automation of innovation or semi-automation of innovation. Innovation studies are not an exemption. We are now at a turning point to reflect on what we have achieved, who we are, and who we will become.

On the terrain of business and organizational processes, automation seems to constitute a standard: if their mindset is suitable, organizational actors will have the tendency to use emergent technologies in those processes that are technology-sensitive (e.g. documentation, transport, recruitment, contracts, sells, services, production, military, medical, legal etc.) (Rodrigo and Palacios 2021). Engaging new technologies in organizing becomes a natural consequence of their social rampancy. When employed in systematic process management (BPI, *Business Process Improvement* (Harrington 1991, Dumas et al. 2018)), automation seriously boosts and optimizes the workflow— allows us to save tangible resources and/or increase productivity. It is, therefore, of no surprise that automation prepares managerial and organizational practice for strategic change and rebuilds it towards innovation.

In this context, our concern is how a pile of existing literature of innovation studies has contributed to the realization of innovation process, and its, at least partial, automation. A huge effort has been devoted to understand and interpret past innovation in a variety of sectors like machines and robotics (Kumaresan and Miyazaki 1999, Pellicciari et al. 2015, Kong et al. 2017), semiconductors and energies (Song and Oh 2015, MacKerron 1995), automotive industry (Llopis-Albert, Rubio, and Valero 2021), information and communication (Bygstad 2010), and how these innovations impact our daily lives and society. Key concepts derived from those stories such as national innovation system, open innovation, academia-industry collaboration, path-dependence, transition or change management, and others that could make this list much longer, have definitely facilitated our understanding of innovation process. If we have enough knowledge on innovation and innovation process, it appears straightforward to consider innovation as a process that can be automated to a significant degree one the basis of socially available resources. However, it seems that pieces of extant achievements are scattered in an unstructured manner. This situation calls for a more systematic approach.

## 2.1. The analogy

Historical development of chemical engineering may serve as a good starting point. Now, most of chemicals include nanomaterials are manufactured by chemical plants where most of operations are automated, and plants themselves are designed with the support of a variety of tools like computer-aided design, computational fluid dynamics, kinetic modeling, and process simulation. But it does not mean that such an automation and computational supports are enabled at the beginning of chemical industry. Before chemical engineering was established, operational know-how of production was related with each material and not structured. This appears to be instructive when it comes to the state-of-the-art in innovation studies. Currently, we have many reports on each innovation which add incremental value to continuous efforts to understand innovation, but they are still highly unstructured. In consequence, they do not significantly improve the big picture of innovations we may have.

In chemical engineering, the breakthrough started with the invention of the concept of 'unit process', which enables detailed quest from input to output in each process and modeling (Groggins 1938). If we can undertake similar endeavor in innovation studies, we can not only obtain a deeper understanding of what innovation process is but also integrate and organize existing pieces of knowledge, model innovation process, and finally—develop the conceptual and empirical equipment for the idea of automation of innovation.

## 2.2. Unit process of innovation

Our understanding of the unit process of innovation builds on the ideas of lifecycle of innovation and the process of innovation which have various accounts in the literature (Tao, Probert, and Phaal 2010, Van de Ven 1986, Utterback 1971). To open the discussion, we simply assume that it captures the following sequential process beginning with observation, through analysis, design, strategic planning, assessment to decision making, and action. In the following, we briefly explain each element of the process (Fig. 1). We will cite only minimum references on the processes, because our aim is to introduce a blueprint of unit process in innovation process but not give a comprehensive literature review.

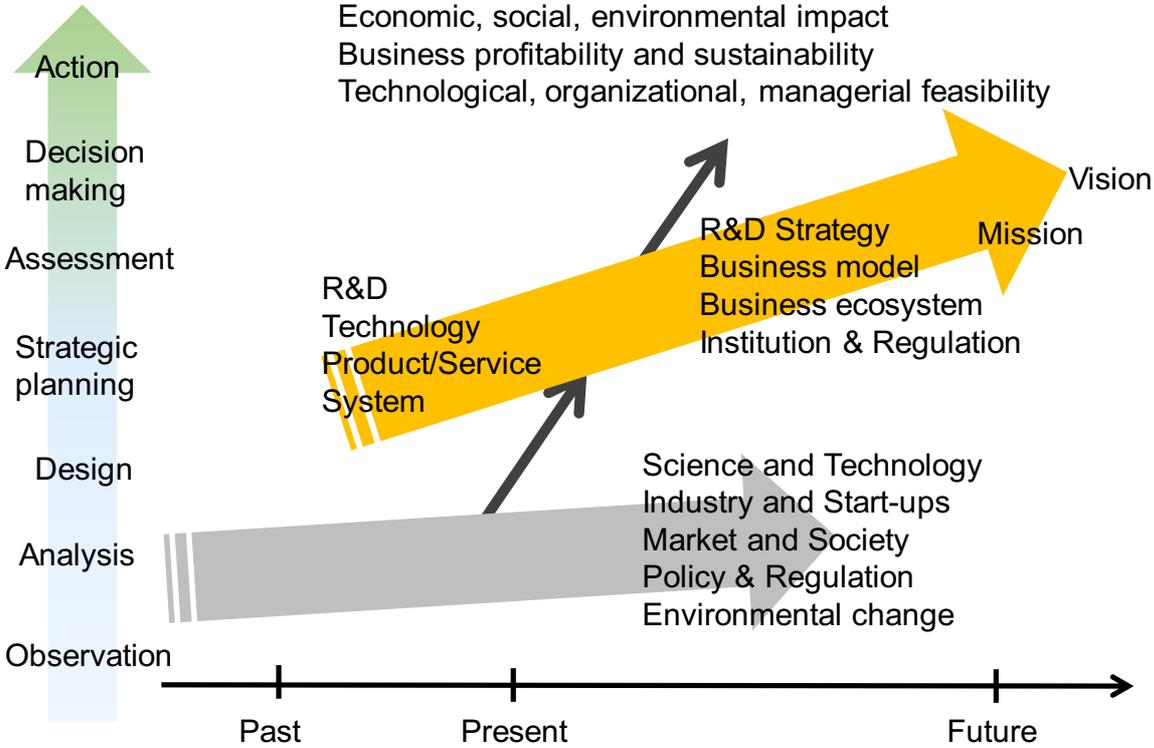

Fig. 1. Unit process in innovation process.

*Observation* includes monitoring of multi-information resources and retrospective data collection. It ranges from science and technology, industry and start-ups, market and society, policy and regulations, and environmental change. Observation based on papers, patents, and industrial data is mature research field in technology and innovation management.

*Analysis* is based on past and present data. One of analyses is forecasting future trend based on existing data and a develop plausible scenario. Another is elucidating hidden mechanisms behind observed phenomena. An example is technological forecasting research based on retrospective patent data.

*Design* does not mean only industrial design of products and services. It includes design of system where products and services are embedded, technology enabling and manufacturing products and services, R&D realizing technology if needed. Examples of design research are axiomatic design (Suh 1990) and design structure matrix (Eppinger and Browning 2012). These are useful frameworks but not devised as a design tool in automation process, and thus we have a much room for further study.

*Strategy and planning* give feasibility that design will embody as business. Strategy is needed at various levels within an organization from R&D strategy, business strategy and business model, corporate strategy, strategy in business ecosystem. Institution and regulation are not always exogenous variables, but they should be treated as endogenous variables in strategic planning process.

*Assessment* of strategy and planning is needed to comprehend technological, organizational, managerial feasibilities, business profitability and sustainability, and economic, social, environmental impacts.

*Decision making* in organization is quite a political issue, reflecting power structure in each organization. If machine learning has more predictive power than experts' judgements, in a naïve sense, decision-making based on AI is beneficial for the organization and society. Strategy research on machine learning and artificial intelligence that started just recently (Balasubramanian et al. 2018) reveals a potential in this area.

It is clear that products and services are offered by a (business) organization. On the figure 1., orange arrow represents those elements that contribute to the process within the organization, while grey arrow represents elements outside it. Science and technology enable them. The role of automation in innovation consists of not only observation and analysis of external events and phenomena but also acceleration of internal process within an organization for product and business development. As discussed later (sect. 3), it is not limited to intra-organizational process but includes multi-level societal changes.

Categorizing existing literature into unit process of innovation framework will help enhance realization of the automation-driven innovation management. Further work is still needed to verify the existence of unit process in innovation, explore implicit unit process, clarify and model each unit process, develop automation tools, and integrate those. Such an endeavor will also enable researchers to notice unresolved issues in each innovation process and give them opportunity to reflect on the usefulness and reliability of their own research in the engineering context of innovation research toward automation. Although investigations into the process of innovation can be curiosity-driven basic research focused on better understanding of innovation, its goal must ultimately be practical. Hence, we intend to highlight the explicitly practical implications, which—in accordance to the idea of BPI—aim to improve organizational practice in the context of innovation.

## 3. HUMAN AND SOCIETAL ASPECTS OF AUTOMATION

A successful organization and integration of the literature on unit process and innovation gives a chance to develop a systematized model which shows the way how automation of innovation is generally possible. However, there is an essential and rather evident factor which differentiates automation in the aforementioned example of chemical engineering and automation of innovation process. The former has exclusive enclosure in chemical plants and sides to avoid unintentional interference of humans. But in the latter, it inevitably involves human actors, organizational, contextual and societal factors in the loop (Greenhalgh et al. 2004, Robertson, Swan, and Newell 1996, Leonidou et al. 2018), because innovation is not merely an invention of artifacts but the implementation of those in society.

Human actors in the social context play interactive roles and affect one another in the circulation of innovations. It is known that those roles may sometimes retard the spread of innovations (Ferlie et al. 2005, Rapport et al. 2018), but when technology and the AI enter the scene, those roles need not generate such challenges (Haefner et al. 2021). Technological innovations are rife in everyday human practice to a significant extent, so they become a reality also in organizations (if people use them in private lives, they tend to use them as organizational actors, too). Technology-infused reality drives organizational change. The research on flexible automation and on the intensification of innovation—the themes known for over two decades—confirms this observation (Cainarca, Colombo, and Mariotti 1989, Bucklin, Lehmann, and Little 1998, Sanchez 1995, Azani and Khorramshahgol 1991, Dodgson, Gann, and Salter 2002).

### 3.1.    Multi-scale impact

Dissemination of innovation not only changes our material life, but also catalyzes the change of our cognitive abilities (Greve and Taylor 2000). And automation, in turn, affects direction of future innovations.

There are two main levels on which this happens:

(1) *cognitive micro-level*: automation saves resources and frees up time which opens new ways to innovate for organizational actors—the more resources an organization has, the better prospects for creative problem-solving and management innovation (on the psycho-cognitive level, this phenomenon has been confirmed empirically (Leszczynski et al. 2017)). Such cognitive transformation has an impact not only on human activities and capabilities but also on development of automation technologies and human-automation interface itself (Jipp 2015, Hamid, Smith, and Barzanji 2017).

(2) *organizational and societal macro-level*: direct transformation of management practice, e.g. automation changes organizational routines and this poses challenges with respect to organizational mindfulness (Kudesia 2019, Levinthal and Rerup 2006) and requires strategic management innovation—departure from accepted management practices and principles. To create value and maintain competitive advantage organizations must use their macro-capabilities for strategic innovation and play a different game (redefinition and restructuration) (Means and Faulkner 2000, Markides 1997). Automation system can be independent and closed, but it may also be an agent interacting with society. Hence, the design of computer-aided systems and ADMs (*Automated Decision-Making* systems (Davenport and Harris 2005, Wachter, Mittelstadt, and Floridi 2017)) that result from such interaction has to take the behavior of other agents into consideration. In consequence, such versions of automation transform strategic planning as well as the ways humans interact and communicate. The increased trust in automation only facilitates this process (Rezvani et al. 2016, Hoffman et al. 2013).

The two levels of influence of automation on managerial and organizational practice and strategy have various implications regarding business, organizational, and societal outcomes. One of them is further development of innovative technologies in a sustainable manner. Strategic innovation has the tendency to support and accelerate technological innovation: to maintain competitive advantage, technologically boosted forms of organizing and management tend to reinforce expansion and further development of emergent technologies.

Naturally, the impact of automation on diffusion of innovations is not—by itself—positive or beneficial as a social transformation (Boyd and Holton 2017), although it is expected to generate economic profits. The emergence of responsible innovation (RI) research (de Saille 2015, Scherer and Voegtlin 2020) reveals that organizational democracy and deliberative engagement should build the social-political framework that secures the efficiency and legitimacy of this transformation (Scherer and Voegtlin 2020, Schneider and Scherer 2019, Brand and Blok 2019). Still, these aspects of innovation are not something that potentially contradicts the idea of automation of innovation. Rather, with respect to the societal nature of the process as a whole, they should be perceived as its necessary elements that set boundary conditions on the level of what is socially and politically acceptable or needed.

In sum, innovation process inevitably involves typically human, organizational, and social dynamics. And automation-driven transformations of managerial and organizational practice are entangled in broad micro-level (psycho-cognitive) and macro-level (organizational) changes that affect those dynamics. The dynamics themselves are subject to democratic regulations.

## 4. TOWARD ACCELERATION OF INNOVATION CYCLE

The above interactive and recursive nature of automation in innovation opens a novel and alternative view on innovation process. To understand the way how the process of innovation is stimulated by the societal and technological factors related to automation, we propose to understand it as a part of the broader innovation cycle. The model of unit process of innovation (Fig. 1) shows the diachronic sequence of operations needed to strategically innovate—from conception to action. This is how particular innovations are produced. The model allows us to highlight selected issues such as, for example, social pull or technology push to innovate, but it is only a part of the broader picture of the circulation of innovations.

### 4.1. The IAS cycle

The circulation itself implies multi-scale societal changes tied both to transformations in the dynamics of human and organizational practice and to the micro-level psycho-cognitive development (see: sect. 3.1). Let us consider the following three key assumptions for relations between technology, automation and innovation:

(1) innovation ecosystems enable automation through technological innovation: today technological innovations are omnipresent, so, using them in business (e.g. broad use of AI) allows to significantly automatize those organizational processes which are technology-sensitive (automation-based process innovation),

(2) far-reaching automation opens new ways and possibilities for significant strategic and management innovation; automation-induced social change forces strategic invention and implementation of new forms managerial and organizational practice,

(3) catalysis effect: technology- and automation-driven, innovative forms of organizing intertwined with societal changes support and accelerate technological innovation by enhancing innovation ecosystems.

The above three suppositions build the cycle of continuous diffusion of innovation. Let us call it The *Innovation-Automation-Strategy* cycle (henceforth: IAS cycle). It can be depicted as follows:

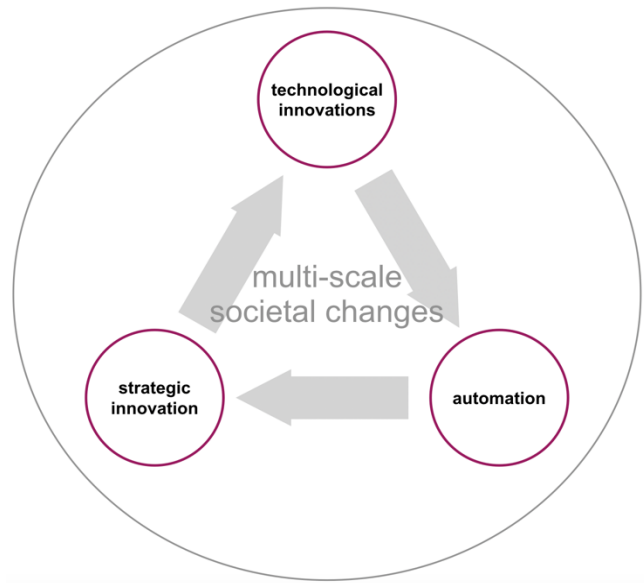

Fig. 2. The IAS cycle

The IAS cycle focuses on selected aspects of technology, automation and strategy that all highlight the interconnections between practical, organizational and technological innovations in a model (or: ideal type (Weber 1903-1917/1949)) of continuous change driven by social changes.

Let us define it in more detail:

**IAS cycle** (definition): a simplified model of multi-scale process in which innovation is profitably transferred from state-of-the-art emergent technologies to technology-sensitive organizational processes where it gives raise to deliberate, strategic management innovation and it further is diffused to new emergent technologies. The whole process is entangled with social change.

The perspective related to the IAS cycle has several consequences among which the following three appear to be of key importance. First, it implies that automation of organizational processes de facto precedes management innovation: emergent technologies first infiltrate organizational structures, practices, and capabilities, change business processes, and then they launch strategic innovation mechanisms. Automation becomes a condition of strategic innovation (not the other way around).

Second, although disruption effects on existent business models, value chains and identities may always appear (Christensen et al. 2018), the IAS cycle is based on the rudimentary assumption that innovations are diffused smoothly (Rogers 2003). In other words, innovations are socially and economically beneficial at each stage of the cycle to the extent that any instances of disruption do not nullify the circulation of automation. If barriers in the adoption of innovations appear, they basically do not hamper the possibility of diffusion of automation in the cycle. Similar observations have recently been made concerning the challenge of adoption of emerging technologies in the conditions of institutional instability (Bonnín Roca et al. 2021). Sleekness is achievable due to typically societal character of the process—if it is realized in the framework of RI and organizational democracy (see: sect. 3.1), diffusion smoothness may be perceived as an intrinsic feature of the cycle. Besides Roger's classic observations (Rogers 2003), various contemporary studies, e.g. in energy industry (Dieperink, Brand, and Vermeulen 2004), agriculture (Hansen 2015) or radio industry (Rossman 2015) confirm the above assumption.

Third, if we accept the view that organizations should strategically support innovations ('the culture of innovation'), then the IAS cycle reveals that automation is no less important in the process of innovation diffusion than deliberate 'managing' of innovations on the micro-level organizational practice. In this sense, automation not only enhances innovations, but also changes the perspective on innovation management (see also sect. 5.5).

The three propositions show that the IAS cycle allows us to holistically capture the place and role of automation for sustainable development: technological innovations change organizational functioning in the way that both boosts technology and may be profitable for organizations and society (Fosso Wamba et al. 2021). Although the IAS cycle does not support the view that innovation process can be automated fully or mechanically, in the sense of the elimination of human control and intervention, it does give reasons for the view that at a certain level of technological development, the process of innovation can be semi-automated as a stable, multi-scale social-organizational AI-infiltrated sequence of actions, interventions and events.

### 4.2. Antecedents of the IAS cycle

The IAS cycle may be perceived as a systematic, organizational consequence of the role and impact of automation (sect. 3) on the shape of the unit process of innovation that emerges from numerous studies of innovations (sect. 2.2). In this sense, it integrates and continues three large streams of research: innovations (including diffusion of innovations), strategy and automation.

The ideas behind this conceptual perspective have been present in the social sciences for a long time, but due to their either early-stage or fragmentary character, they could not open similar vistas and research questions. The early works on automation (Simon 1966, Bright 1958) may be perceived as correctly forecasting the roles of automation in organizational and industrial contexts. Especially, the tradition of the Carnegie School (Simon 1969, March and Simon 1958, Cyert and March 1963), highlighting bounded resources of organizational actors (including managers) and the need to rely on programming behavior appears as important. When related to strategic decision-making (Eisenhardt and Zbaracki 1992), computer-aided decision systems and contemporary views on automation, it opened a new perspective on organizational processes. Currently, it has been colonized by the AI- and machine learning-related research on automation (Balasubramanian, Ye, and Xu 2020), which in some ways revives and continues ideas from the Carnegie School.

Another root of the IAS cycle is the early work on diffusion of innovations (Katz, Levin, and Hamilton 1963), and later—its integrative models (MacVaugh and Schiavone 2010). Contemporary studies of the impact of automation on innovations as well as of the complementarity between business process management and digital innovation (Mendling, Pentland, and Recker 2020) play equally important roles as pillars of knowledge behind the cycle in which innovations are entangled.

# 5. IMPLICATIONS FOR FUTURE RESEARCH AND LIMITATIONS

In the previous section, we briefly considered which ideas and streams of research support and build the foundation of the IAS cycle. Now we will examine key practical implications of the cycle and briefly discuss areas of future research related to those implications. IAS is entangled in the whole variety of complex issues the knowledge of which is currently at an early stage (Dwivedi et al. 2019, Haefner et al. 2021). More systematic investigations of those issues will allow us to shine a new light on the vast array of questions, open research opportunities, sketch a research agenda and understand its limitations. Below, we group those issues under five main headings: The first three of them refer to the elements of the cycle, the fourth deals with its social transformation context and the fifth raises key conceptual implications. Finally, the questions of key limitations are discussed.

## 5.1. Technological Innovation and Automation

One of the most interesting implications of the proposed perspective on innovations is how emerging technologies and AI may directly contribute to the acceleration of the innovation cycle and how they facilitate the social automation of innovation. Currently, those roles of new technologies are rather implicit and context-dependent, and little known (cf. Satchell 2020). In consequence, some technological inventions may effectively push automation in a given environment, at the same time facing difficulties to play this role in another environmental context (Bonnín Roca et al. 2021). Contextual turbulences and institutional dynamics may also significantly limit the adoption of the IAS cycle.

This observation also opens paths to explore important and related question of tools automating innovations and facilitating the whole cycle in which they are entangled. This is a task not only for technological innovation design (Norman and Verganti 2014, Verganti 2009), but also for development and implementation science.

Additional, highly important question for the relation between technological innovation and automation is the possibility of measuring and designing the acceleration of innovations in the cycle. Machine learning approaches promise a good way forward. Recent studies in bioinformatics and drug discovery (Réda, Kaufmann, and Delahaye-Duriez 2020, Xia 2017), business partner recommendation systems (Mori et al. 2012), or even automated methods of discovering novel research targets in science (Ogawa and Kajikawa 2017) give interesting results on the micro-scale. Thus, sufficient accumulation of such studies may generate interesting macro-level outcome for understanding the acceleration in the cycle.

## 5.2. Automation and Strategic Innovation

According to one of our assumptions, automation opens new ways and possibilities for significant strategic innovation (sect.: 4.1.). Currently, our knowledge how automation and AI drive organizational and strategic innovation and how organizations exploit automation is limited, mainly because it focuses on technical, ergonomics-related aspects of its implementation (Dwivedi et al. 2019, Wickens et al. 2015, Parasuraman 2000). In consequence, we do not know much about which organizational paths are available to adopt automation in the way that it allows for management practices to enter the loop according to IAS. In a similar vein, the study of psycho-cognitive dimension of strategic innovation that allows managers and organizations to facilitate the automation cycle (technology-mindset interaction (Ringberg, Reihlen, and Rydén 2019)) is also at an early stage.

On the other hand, assuming that organizations widely adopt the IAS cycle to accelerate innovations, we will still have interesting paths to study the way how they differentiate their strategies and how they should orchestrate stakeholders and other agents. One pole of scenarios is to delegate our decision to AI, which is derived from data-driven and model-based analysis including psycho-cognitive dimensions of the others. Another is business-as-usual scenario where managers make decisions by their own judgments and responsibilities. Plausible one is a mixture of those. We design boundaries and rules of strategic games where agent-based simulations are run. We can utilize the simulation results for our decision making and can also delegate our AI-based agent to negotiate with the other agents. In any scenarios, we face the need of different approaches in strategic decision making than we currently make (Balasubramanian, Ye, and Xu 2020), which will open a new frontier both for research and practice.

## 5.3. Strategic Innovation and Technological Innovation

The IAS cycle has various strategic consequences for technological innovations. This means, foremost, initiation of investments in emerging technologies. If new technologies may directly contribute to the acceleration of the innovation cycle (sect. 5.1), then this process has strategic consequences, for example, for development of SNM (Strategic Niche Management research) (Schot and Geels 2008). IAS cycle may help survive and grow grassroot innovations (Hargreaves et al. 2013) or increase their resilience in face of institutional turbulences (Bonnín Roca et al. 2021). From a viewpoint of technological innovation, if these niches are promising both for economic and social aspects, automation process will rationally judge their promising potential and invest on those. Thus, IAS cycle accelerates innovation without a bias on the past development path and strategic inertia. However, on this case, how should we differentiate strategy? And what strategy should be taken by actors on the current technological regime?

Although the innovation cycle assumes that innovations are diffused smoothly (sect. 4.1), it does not remove the question of disruption of innovations and struggles related to disruptive shift (Nagy, Schuessler, and Dubinsky 2016, Christensen et al. 2018). However, the question is still open, we stipulate that the IAS cycle offers a new forward-looking perspective which supplements the focus on the impact of innovations and new technologies on existing markets, enterprises and business models which is typical for the analyses of disruptive effects. The extent to which there is an interplay between these perspectives is another avenue of study.

### 5.4.    Multi-scale societal change

The IAS cycle is entangled in multi-level social transformation (sect. 4.1), which engages it into typical questions in the area of AI and innovation studies. For example, it may help better understand and develop the concept of innovation ecosystem (Tsujimoto et al. 2018, Granstrand and Holgersson 2019). Further, it may contribute to expansion of our knowledge that currently results in redefinitions of human work. More specifically, the understanding of such questions as labor displacement (Gruetzemacher, Paradice, and Lee 2020), augmentation (Raisch and Krakowski 2020) or human-AI symbiosis (Jarrahi 2018) may obtain additional support under the umbrella of the IAS cycle. Currently, the impact of AI on human work is rarely associated with deepened knowledge of management innovation (which is basically interpreted through institutional, cultural, fashion-related or rationalistic lenses (Birkinshaw, Hamel, and Mol 2008). A simple example is promotion system in an organization. After IAS cycle is implemented, how to assess, evaluate, and valuate achievements and contributions of each worker? In what ways, does new schema of promotion system affect workers' tasks, motivations, and incentives? How should we face those situations? It is clear that we have societal challenges at micro, meso, and macro-levels.

Analogically to the problem of disruption (sect. 5.3), the innovation cycle is involved in various regulation-related challenges. We accepted the idea that organizational democracy and the framework of RI (sect. 3.1) belong to the IAS cycle, but this maneuver does not close the discussion related to the practical limitations of RI (de Hoop, Pols, and Romijn 2016). Other social challenges such as the AI ethics (Hagendorff 2020, Jobin, Ienca, and Vayena 2019), good AI society (Fosso Wamba et al. 2021) or sustainable innovation (Cillo et al. 2019) strengthen the context of macro-level regulation. Recent studies (Silvestre and Țîrcă 2019, Bonnín Roca et al. 2021) show that the sustainability trajectory may not be a problem of IAS cycle. These contexts request to implement regulatory mechanisms in IAS cycle. In this respect, we stipulate that the cycle has the potential to open a novel approach to sustainable and responsible innovation.

The social transformation behind the cycle of innovation has an impact on the micro-level of human actors and their psychology. By acknowledging this micro-level (sect. 3.1), the innovation cycle may contribute to development of conceptual models of technology in organizational psychology (Morelli et al. 2017, Tonidandel, King, and Cortina 2016), improve our understanding of the impact of big data and AI on psychology in industrial contexts, and thus help face the problem that new technologies frequently outpace human needs. IAS cycle may also expand our understanding how micro-level strategic innovations of individual actors and their interactions contribute to the innovation cycle.

Another newly emerging grand social challenge is how the IAS cycle deals with such macro-level causes of automation acceleration as pandemics (Chernoff and Warman 2020, Coombs 2020, Brem, Viardot, and Nylund 2021). On our view, the framework we propose may not only help better understand such processes but also facilitate organizational change related to them and contribute to reconstruction of global governance system.

## 5.5.    Conceptual transformations

One of the most vividly discussed societal problems related to automation—its impact on labor market (sect. 2)—has palpable conceptual dimension: it leads to the revision of concepts of work and job (McKenna 2017, Schoukens and Barrio 2017). Such revisions are not the only ones with which the IAS cycle is intertwined.

Interesting reconceptualizations embrace the concept of innovation management. Understanding automation within the IAS cycle changes the perspective on innovation management (sect. 4.1). The need of such transformations has already been observed (Volberda, Van Den Bosch, and Heij 2013). Due to the increased role of automation and digitalization, the standard idea of 'managing innovation' (Bessant and Tidd 2013) changed its sense—despite some previous opinions (McCabe 2002), innovation is not anymore reproduction of the past. In this context, the idea of managing as designing (Boland et al. 2008) is increasingly more apt and it strengthens the role of top management (Nell et al. 2020).

Finally, the challenge of conceptual transformation refers also to the very concept of innovation: the understanding of innovation as something that is simply introduced by rational (deliberately "innovating") individuals or entrepreneurs (Drucker 2015) may not be fully appropriate in the framework of the IAS cycle. The extent of revision is a matter of future studies, however. At any rate, the proposed model should facilitate also such theoretical and conceptual transformations.

## 5.6.    Innovation studies

The perspective and model of innovation build on several ideas and streams of research some of which have already been explored in innovation studies (sect. 4.2). Still, the question of the extent to which we should depend on past achievements of innovation studies for innovation automation remains open. Future studies will also determine the scope of processes to which innovation studies can contribute to and what is the most effective (IAS cycle-enhancing) style of output of studies, papers, structured knowledge, or tools. In these respects, significant, automation-driven changes are possible and they may embrace also the research process of innovation studies itself. Limitations in this area seem to be defined only by the extent to which machine learning and the AI cannot infiltrate research.

The above constraint takes us to the final, indeed crucial, observation that pertains to all the above implications and research prospects for the IAS cycle. We should be aware that possible developmental scenarios for the cycle of innovation will look differently depending on the degree to which particular innovation processes are merged in cyber, physical and social worlds. Although machine learning and AI infiltrate many organizational and business processes that take place in a natural and social world, it is still the cyber world where they generate controllable scenarios. Successful implementation of those scenarios in a physical and social world depends on various factors.

## 6. CONCLUSION

The IAS cycle gives a chance to unlock an integrated, comprehensive, and intensified approach to relations between technological innovations, process management and management innovation. These areas and knowledge about them are mature enough to reveal their interdependencies and open a whole new perspective on how we—scholars, technology developers, organizational actors, managers and members of global society—use the AI-driven new technologies. The IAS cycle introduced in this paper can help make precise and adjust the idea of innovation management to the demands and pressures which developing technologies make on managerial and organizational practices. This involves thorough reexamination of extant knowledge in innovation studies, good understanding of the role of automation and readiness to draw deep consequences of its impact on innovation management.

The innovation cycle introduced in this paper presents the elements of process of technological innovation as entangled in social and technological transformation. We showed that it is a subject to a peculiar automation due to the omnipresence of new technologies. On that basis, we proposed to make complex practical findings for innovation management and strategy. The ultimate goal of introducing the IAS cycle is strengthening innovation studies.